\documentstyle[preprint,prb,aps,tighten]{revtex}
\preprint{HKBU-CNS-9830}
\begin{document}
\draft
\title{Disturbance spreading in incommensurate and quasiperiodic
systems}

\author{Bambi\ Hu$^{1,2}$, Baowen Li$^{1,3}$\footnote{Author to whom 
correspondence should be addressed. E-mail: bwli@phibm.hkbu.edu.hk}, and
Peiqing\ 
Tong$^{1,4,5}$} 
\address{
$^{1}$Department of Physics and Centre of Nonlinear Studies, Hong Kong 
Baptist University, Hong Kong, China,\\ 
$^{2}$Department of Physics, University of Houston, Houston, Texas 77204, 
USA\\
$^{3}$ Department of Physics, National University of Singapore, 119260 
Singapore\\
$^{4}$CCAST(WORLD LABORATORY) P.O. Box 8730, Beijing, 100080, China\\
$^{5}$Department of Physics, Nanjing Normal University, Nanjing, Jiangsu, 
210097, China}

\date{\today}
\maketitle
\begin{abstract}

The propagation of an initially localized excitation in 1D
incommensurate, quasiperiodic and random
systems
is investigated numerically. It is discovered that the time evolution of
variances $\sigma^2(t)$ of atom displacements depends on the initial
condition. For the initial condition with nonzero momentum, $\sigma^2(t)$
goes as $t^\alpha$ with $\alpha=1$ and $0$ for incommensurate
Frenkel-Kontorova (FK) model at $V$ below and above $V_c$ respectively;
and $\alpha=1$ for uniform, quasiperiodic and random chains. It is also found
that $\alpha=1-\beta$ with $\beta$ the exponent of distribution
function of frequency at zero frequency, i.e., $\rho(\omega)\sim
\omega^{\beta}$ (as $\omega\rightarrow 0$).  For the initial condition
with zero momentum, $\alpha=0$ for all systems studied. The underlying 
physical meaning of this diffusive behavior is discussed. 

\end{abstract}

\pacs{05.60.Cd, 61.44.Fw, 61.44.Br, 63.90.+t}


\section{Introduction}

The fast development of nanotechnology makes one-dimensional (1D) systems 
like quantum wire\cite{WWUL85} and nanotubes\cite{Nanotube} available in 
laboratories nowadays. The study of physical properties such as  
localization, thermal conductivity, and electron conductance, etc., in
these 
systems become more and more important\cite{Conduct}. In the 
past years, many works have been done on 
the waves (classical and quantum) propagation and localization 
in 1D disordered media. Particular interests are paid to  
propagation of the classical waves in random media (see e.g. 
\cite{Sheng90}), and electron transport in 1D 
disordered solids \cite{ETP,Lif88}. 
In comparison with the disordered systems, much less is studied
about the transport and diffusion properties of incommensurate and 
quasiperiodic systems, even though the incommensurate structures appear in many 
physical systems such as
quasicrystals, two-dimensional electron systems, magnetic superlattices,
charge-density waves, organic conductors, and various atomic monolayers
absorbed on crystalline substrates\cite{Incomm}. 

In this paper, we would like to study the classical transport of an
initially localized excitation in 1D
incommensurate, quasiperiodic and random systems so as to better 
understand
transport processes  and 
relaxations
properties in these systems. Recently, quantum
diffusion in two families of 1D models have attracted
much attention\cite{Geisel,ar92,wi94}.  Typical examples are
the kicked rotator and kicked Harper models from the field of quantum chaos,
and the  Harper model and the tight-binding model associated with
quasiperiodic sequences. Many interesting dynamical behaviors, such as
quantum localization and anomalous diffusion, and their relationships with
energy spectra have been investigated in these
systems\cite{Geisel,ar92,wi94}.  
However, the classical transport in incommensurate and quasiperiodic 
sytems and its relation with
phonon frequency distribution, as well as the diffusive behavior
dependence on 
initial condition, etc. have not yet fully investigated up to now.

The information of disturbance spreading in such system reflects the
interior structures of the underlying system. As we shall see later,
the spreading properties are determined largely by the density of states,
in particular by the phonon model near the zero frequency. 

\section{Models and numerical results}

\subsection{Incommensurate chain} 
The Frenkel-Kontorova (FK)
model\cite{fr38} is invoked as a prototype of an incommensurate 
chain in this paper. This model is a 1D atom chain
with an elastic nearest neighbor interaction and
subjected to an external periodic potential.  
Most works on this 
model in the past two decades have been concentrated on ground state 
properties and phonon spectra, etc.
The 1D FK model is described by a dimensionless 
Hamiltonian 

\begin{equation}
H=\sum_{n}\left[\frac{p^2_n}{2}+\frac{1}{2}(x_{n+1}-x_{n}-a)^2-V\cos(x_{n}) 
\right],
\end{equation}
were $p_n$ and $x_n$ are  momentum and position of the $n$th atom,
respectively. $V$ is the coupling constant,
and $a$ is the
distance between consecutive atoms without external potential. Aubry and
Le Da\"eron\cite{AubryLeD} showed that the ground state configuration is
commensurate when $a/2\pi$ is rational and 
incommensurate when $a/2\pi$ is irrational. For an incommensurate 
ground state, there are two different configurations
separated by the so-called {\it transition by breaking of analyticity}  
predicted by Aubry\cite{Aubry}. This transition survives the quantum
fluctuation\cite{hl98}. Moreover, in contrast to other 1D nonintegrable 
systems such as the Fermi-Pasta-Ulam chain\cite{FPU}, the FK chain shows a 
normal thermal conductivity \cite{HLZ98}. For each irrational number $a/2\pi$ 
there exists a critical value $V_c$ separating the sliding state 
($V<V_c$) from pinned state ($V>V_c$). The
$V_c=0.9716354\cdots$ corresponds to the most irrational number, golden
mean value $a/2\pi=(\sqrt{5}-1)/2$. Without loss of generality, we
restrict ourselves to this particular value of $a$ in the numerical
calculations throughout the paper, and it
is approximated by a converging
series of truncated fraction: $F_n/F_{n+1}$, where
$\{ F_n\}$ is the Fibonacci sequence.
 
The equation of motion for the $n$th atom in the FK model around
its equilibrium position is

\begin{equation}
\frac{d^2\psi_n}{dt^2}=\psi_{n+1}+\psi_{n-1}-[2+V\cos(x_n^0)]\psi_n,
\label{eq:dyn}
\end{equation}
where $x_n^0$ is the equilibrium position of the $n$th atom  at ground 
state, and $\psi_n$ is the normalized displacement from the equilibrium 
position. In fact, to obtain Eq.(\ref{eq:dyn}), we have written the whole 
displacement of particle as $x_n = x_n^0 + \epsilon \psi_n$, where 
$\epsilon (\ll 1)$ is a small parameter.
To quantify the disturbance spreading, the 
variance of displacements

\begin{equation}
\sigma^2(t)=\frac{1}{N}\sum_{n=1}^{N} | \psi_n(t)-\psi_n(0) |^2
\end{equation}
is calculated
by two numerical methods. The first one is the Runge-Kutta 
method of the fourth  order to integrate Eq. 
(\ref{eq:dyn}) for a given initial condition with free boundary.  
The second one is to find eigenfrequencies 
$\omega_j$ and eigenvectors $\alpha_n(j)$ of equation
\begin{equation}
-\omega^2\psi_n=\psi_{n+1}+\psi_{n-1}-[2+V\cos(x_n^0)]\psi_n.
\label{eq:eig}
\end{equation}   
The solution of Eq. (\ref{eq:eig}) can be expressed in the 
following form:
\begin{equation}
\psi_n(t)=\sum_{j=1}^{N} \left[ A_j\cos(\omega_j t)+B_j\sin(\omega_j 
t)\right] \alpha_n(j) \label{eq:sol}
\end{equation}
where the coefficients $A_j$ and $B_j$ are determined by initial
conditions. Contrasting to the quantum diffusion, the classical evolution 
Eq. (\ref{eq:dyn}) is of the second order of derivative.  
Thus initial conditions for $\psi_n$ and $d\psi_n/dt$ are needed. One of
our main 
findings is that the spreading behavior depends on the initial 
condition. For the initial condition  
$$ \psi_n=0\,\,\, \mbox{and}\,\,\, d\psi_n/dt=\delta_{n,n_0}$$ 
which is called type I, we have 
$\sigma^2\sim t^{\alpha}$, and $\alpha$ is equal to $1$ and $0$ for
$V<V_c$ and $V>V_c$, respectively. For the initial condition 
$$\psi_n=\delta_{n,n_0}\,\,\,\mbox{and}\,\,\, d\psi_n/dt=0$$ 
which is called  type II, $\sigma^2\sim 
t^0$ for any $V$. (Of course, there is another type of initial condition, 
i.e., $\psi_n=\delta_{n,n_0}$ and $d\psi_n/dt =\delta_{n,n_0}$. Our 
numerical calculations show that the spreading behavior in this case 
is the same as that of type-I initial condition.)
Figure \ref{fig1} shows the
typical time evolution $\sigma^2(t)$ for the FK model.
In numerical calculations, we first
obtain the ground state positions of $N$ atoms in the FK chain
by the gradient method for free boundary, i.e., $x_0\equiv0$ and
$x_N=Na$.  The results of Fig. \ref{fig1} are obtained by
the integration method for $N=F_{19}=10946$. $\sigma^2(t)$ is also
calculated
by the second numerical method for the FK chains of small size,
which gives rise
to the 
same results. 

It is worth pointing out that 
the above-mentioned results valid only for the evolution time less than 
a critical value $t^*$, where $t^{*} \sim
N/2v$ and $v$ is the velocity of sound. For our FK model, $v\approx 1$.
After this critical time, i.e., $t>t^{*}$, the
power relation of $\sigma^2(t) \sim t^\alpha$ is destroyed due to 
the finite size effect. 

To get a clear picture of the spreading of a disturbance in an
incommensurate chain
with two different initial conditions, we plot $\psi_n(t)$ in Fig. 2 for
the FK chain of $V=0.4$.  The intensity of gray scale represents the 
amplitude of the displacement of the particle. Because of the huge 
amount of the data, we record $\psi_n(t)$ at a time interval of 20 time 
steps, which leads to some discontinuity. Figure 2(a) demonstrates the 
evolution with initial 
condition $\psi_n =0$ and $d\psi_n/dt =\delta_{n,n_0}$, and Fig. 2(b) 
shows that with initial condition $\psi_n=\delta_{n,n_0}$ and $d\psi_n/dt 
=0$. The difference is clear. In the later case the disturbance spreads 
out in both direction, and the particle remains almost at rest after 
the disturbance passes it. However, in the first case, wherever the 
disturbance spreads, the particle will be excited and keeps moving. 
In the cantorus regime ($V>V_c$), the disturbance spreading in the FK chain 
is similar to the case in Fig. 2(b) regardless of the initial condition.

\subsection{Quasiperiodic and random chains} 

We turn now to study 
of disturbance spreading in uniform, 
quasiperiodic, and random chains.  The equation of motion 
can be written as 
\begin{equation}
\frac{d^2\psi_n(t)}{dt^2}=k_n\psi_{n+1}+k_{n-1}\psi_{n-1}-(k_n+k_{n-1})\psi_n.
\label{eq:uni}
\end{equation}
If $k_n=k$ for all $n$, it corresponds to a uniform chain. For quasiperiodic 
chains, $k_n$ takes two values $k_1$ and $k_2$ which are arranged 
according to some deterministic quasiperiodic substitute rules\cite{Aperiodic}. 
Here  we discuss four types of quasiperiodic
chains. They are Fibonacci, Thue-Morse, Rudin-Shapiro, and period-doubling 
chains, respectively. The substitute rules for them are:
$k_1\rightarrow k_1k_2, k_2\rightarrow k_1$ (Fibonacci); 
$k_1\rightarrow k_1k_2, k_2\rightarrow k_2k_1$ (Thue-Morse); 
$k_1k_1\rightarrow k_1k_1k_1k_2, k_1k_2\rightarrow k_1k_1k_2k_1,  
k_2k_1\rightarrow k_2k_2k_1k_2, k_2k_2\rightarrow k_2k_2k_2k_1$ 
(Rudin-Shapiro);  
$k_1\rightarrow k_1k_2, k_2\rightarrow k_1k_1$ (period-doubling).
According to the classification based on the eigenvalues of generating
matrix defined by Luck\cite{Luck}, they are bounded (Fibonacci,
Thue-Morse), unbounded (Rudin-Shapiro), and marginal (period-doubling).
For
comparison,  $\sigma^2(t)$ for random chain is also studied.
In this case, the values of $k_n$ are taken $k_1$ and $k_2$ with 
the same probability. 
Figure \ref{fig3} shows typical time evolutions of variances for four
quasiperiodic and random chains. The disturbance spreading behaviors in
these chains are the same as that of the incommensurate FK model at
$V<V_c$, namely, $\sigma^2(t)\sim t$ for all these 
chains with
the initial condition of nonzero momentum $[d\psi_n(0)/dt\neq 0]$, and
$t^0$ 
for all these chains with initial condition of zero momentum 
$[d\psi_n(0)/dt =0]$. 

\subsection{Relationship with phonon spectrum}

Figures 1-3 are the main results.
They demonstrate that the disturbance spreading 
depends crucially on the 
initial condition. In the following, we would like to understand this 
peculiar behavior in terms of the phonon spectra. 

The coefficients $A_j$ and $B_j$ in Eq. (5) are:
$A_j=0$ and $ B_j=\alpha_{n_0}(j)/\omega_j$ for type-I boundary 
condition; $ A_j=\alpha_{n_0}(j)$ and $B_j=0$ for type II.
Therefore the solutions of Eq. (\ref{eq:dyn}) are
\begin{eqnarray}
\psi_n &=&
\sum_{j=1}^{N}\sin(\omega_j 
t)\alpha_n(j)\alpha_{n_0}(j)/ \omega_j,\qquad \mbox{type I},\nonumber\\
\psi_n &=&
\sum_{j=1}^{N} \cos(\omega_j t)\alpha_n(j)\alpha_{n_0}(j), \qquad 
\mbox{type 
II},
\end{eqnarray} 
respectively. As $N\rightarrow\infty$, we have  
\begin{eqnarray}
\frac{1}{N}\sum_{n=1}^N\psi_n^2 &\sim & \int_{\omega_{min}}^{\omega_{max}}
\sin^2(\omega t)\alpha_{n_0}^2\rho(\omega) d\omega/\omega^2,\qquad 
\mbox{type 
I,}\nonumber\\
\frac{1}{N}\sum_{n=1}^N\psi_n^2 & \sim & 
\int_{\omega_{min}}^{\omega_{max}}\cos^2(\omega t)
\alpha_{n_0}^2\rho(\omega) d\omega \qquad \mbox{type II},
\label{eq:cas12}
\end{eqnarray}
respectively, where $\omega_{max}/\omega_{min}$ is the maximum/minimal 
frequency of phonon spectra and $\rho(\omega)$ is the density of the
phonon 
spectra.

The difference between the integrands in Eq. (\ref{eq:cas12}) for type I 
and type II  lies in the factor $1/\omega^2$. 
As time increases, the dominant contribution of the integral in Eq. 
(\ref{eq:cas12}) for type I comes from the
integrand around $\omega=0$. The integrand for type II is an 
oscillated 
function of time $t$, and so is the integral. Therefore the 
reason for different behaviors of these chains for 
different initial conditions is due to the coefficients $B_j$ in 
Eq. (\ref{eq:sol}). If $B_j$ is equal to
zero, i.e., the initial condition with zero momentum, $\sigma^2(t)$ is 
an oscillated function of time.
If $B_j$ is nonzero, i.e., the initial condition with 
nonzero momentum, $\sigma^2(t)$ is proportional to $t$.

In fact, the integral in Eq. (\ref{eq:cas12}) for type-I boundary
condition can be 
written as 
\begin{equation}
\int_{\omega_{min}t}^{\omega_{max}
t}t \sin^2(\tilde{\omega}) \alpha_{n_0}^2
\rho(\frac{\tilde{\omega}}{t}) d \tilde{\omega}/\tilde{\omega}^2.
\label{int}
\end{equation}
If the
distribution function of frequency has the scaling behavior
$\rho(\omega)\sim \omega^{\beta}$ at 
low frequency ($\omega\rightarrow 0$), then one has $\alpha=1-\beta$.
For the uniform chain, it is well known that
$\rho(\omega)=2/(\pi\sqrt{\omega_m^2-\omega^2})$, thus $\beta=0$. 
We also discover that for the Fibonacci chain and random chain, the 
distribution
function of frequency at low frequency are the same as that of uniform
chain\cite{Lif88,Lu86}. To demonstrate this, we calculate the integrated 
distribution function
of frequency (IDFF) for these quasiperiodic chains by directly diagonalizing
chains of finite length, and plot them in Fig. \ref{fig4} as a function
of $\omega$. The results suggest that
for all these quasiperiodic chains the IDFF is proportional to $\omega$ at low
frequency, thus $\beta=0$, which is the same as that of uniform and random
chains  so that  the relationship $\alpha=1-\beta$ is satisfied for all
these systems. 

For the incommensurate FK model, it is well known that there is a
zero-frequency phonon mode for $V<V_c$, whereas there is a phonon 
gap for $V>V_c$.  From above discussion, we know that the zero-frequency
phonon mode plays a key role in  the time behavior of
$\sigma^2(t)$. The time behavior of $\sigma^2(t)$ in the 
incommensurate FK chain at
$V<V_c$ suggests that the low frequency behavior of distribution function
of frequency is the same as that of those chains discussed above. The curves 
shown in
Fig. \ref{fig4} indeed demonstrate this. 
But for the incommensurate FK chain at $V>V_c$,
$\omega_{min} >0$, thus the integral in Eq.
(\ref{eq:cas12}) for type I is an oscillated function and the time 
behavior of
$\sigma^2(t)$ is also an oscillated function of time.  The case $V=V_c$ is
critical. The phonon spectrum of the FK chain at $V_c$ is different from
that of
$V<V_c$. It has self-similar structure and is point spectrum [see Fig. 
\ref{fig5}(a)]. Therefore
there is no inverse power relation between $\rho(\omega)$ and $\omega$ at
low frequency. It implies that the results would depend on the length of the 
chain in numerical calculation. This is illustrated in Fig. \ref{fig5}(b), 
where we plot $\sigma^2(t)$ as a function of $t$ for 
the FK chains of different length at $V=V_c$. 

\section{Conclusion and discussions}

We have studied the disturbance spreading  in
incommensurate, uniform, quasiperiodic and random chains. We
have found that the time evolution of variance $\sigma^2(t)$ depends on the 
initial conditions. Its behavior is determined by the
density of phonon frequency around zero frequency.
For the initial condition of zero momentum, 
$\sigma^2(t)\sim t^0$ for all kinds of chains studied in this letter. For 
the initial condition of
nonzero momentum,  $\sigma^2(t)\sim t^\alpha$, $\alpha=1$ for uniform,
quasiperiodic, random chains, and incommensurate FK chain at $V<V_c$. 
Although other physical properties differs from system to system, the
time behavior of $\sigma^2(t)$ are the same for all these systems. For
the incommensurate FK chain at $V>V_c$, $\sigma^2(t)$ is an oscillated 
function of
time.  This different behavior of the incommensurate FK chain at different
$V$ regimes might provide us a different approach to detect the 
transition by breaking of analyticity experimentally. \\

This work was supported in part by grants from the Hong Kong 
Research Grants Council (RGC) and the Hong Kong Baptist University Faculty 
Research Grant (FRG). Tong's work was also supported  by Natural 
Science Foundation of Jiangsu Province and Natural Science Foundation of 
Jiangsu Education Committee, PR China.

\begin{figure}
\caption{$\sigma^2(t)$ of the incommensurate 
FK chains with different values of $V$ for different initial conditions. The 
solid and dotted
lines correspond to $V=0.4$ and $V=1.6$, respectively. (a) Initial 
condition $\psi_n(0)=0$ and $d\psi_n(0)/dt=\delta_{n,n_0}$; (b) 
initial condition $\psi=\delta_{n,n_0}$ and $d\psi_n(0)/dt=0$. 
The length of the FK chain is $N=10946$. Initial excitation locates at 
the center of the chain, namely, $n_0=N/2+1$.} 
\label{fig1}
\end{figure}

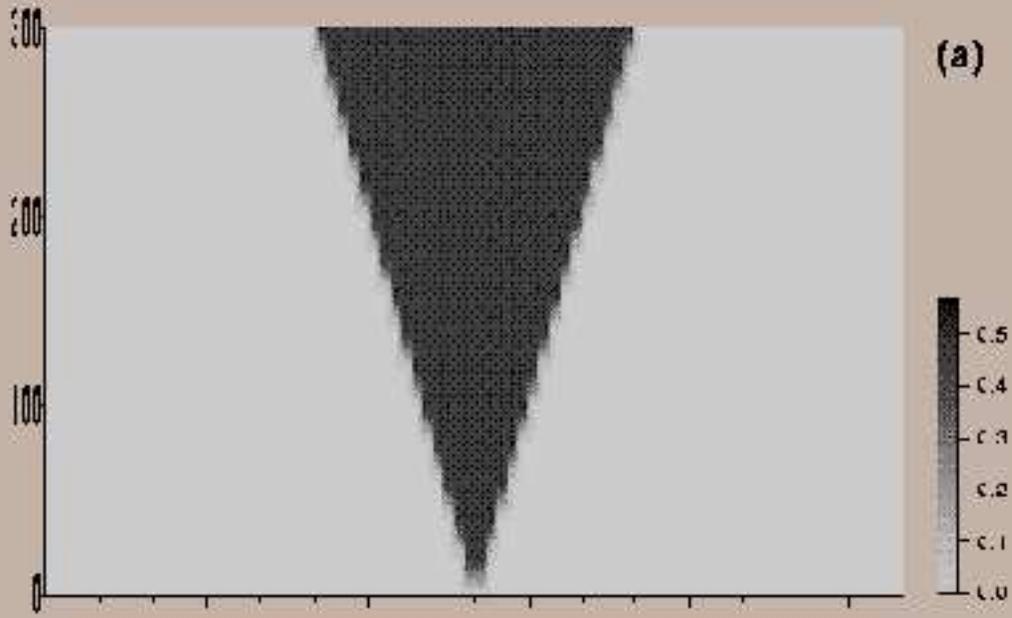
\begin{figure}
\caption{Time evolution of the displacement $\psi_n(t)$ for an
incommensurate FK chain of $V=0.4$ with different initial conditions: (a)  
$\psi_n(0)=0$ and $d\psi_n(0)/dt=\delta_{n,n_0}$; (b)
$\psi=\delta_{n,n_0}$ and $d\psi_n(0)/dt=0$. The length of the FK chain 
is $N=1597$. The initial excitation locates at the middle of the chain, 
i.e., $n_0= 799$.} 
\label{fig2}
\end{figure}

\begin{figure} 
\caption{$\sigma^2(t)$ of quasiperiodic and random 
chains with different initial conditions. 
The solid, dotted, dashed, long dashed, and 
dot-dashed lines correspond to Fibonacci, Thue-Morse, period-doubling, 
Rudin-Shapiro, and random chains, respectively. (a) Initial 
condition $\psi_n(0)=0$ and $d\psi_n(0)/dt=\delta_{n,n_0}$; (b) 
initial condition $\psi=\delta_{n,n_0}$ and $d\psi_n(0)/dt=0$. 
The chain length is $N=10946$ for the Fibonacci chain and $N=8192$ 
for the Thue-Morse, period-doubling, Rudin-Shapiro, and  random chains. }
\label{fig3}
\end{figure}

\begin{figure}
\caption{The integrated distribution function of  frequency (IDFF) as 
a function of $\omega$ at low frequency for different chains. 
The solid, dotted, dashed, long dashed, and 
dot-dashed lines correspond to the Fibonacci, Thue-Morse, 
period-doubling, Rudin-Shapiro chains,
and the incommensurate FK chain at $V<V_c$, respectively.  The results are 
obtained by directly
diagonalizing chains of finite length. The chain length is
$N=1597$  for the  Fibonacci and the FK chain, and $N=2048$ for the 
Thue-Morse, period-doubling, and Rudin-Shapiro chains.} 
\label{fig4} 
\end{figure}

\begin{figure}
\caption{(a) The distribution function of frequency $D$ as a function of 
$\omega$ at low frequency for the incommensurate FK at $V=V_c$. 
(b) the time evolution of variance $\sigma^2(t)$ in the incommensurate 
FK chain at $V=V_c$ for different lengths. 
The initial 
condition is $\psi_n(0)=0$ and $d\psi_n(0)/dt=\delta_{n,n_0}$. }
\label{fig5} 
\end{figure}

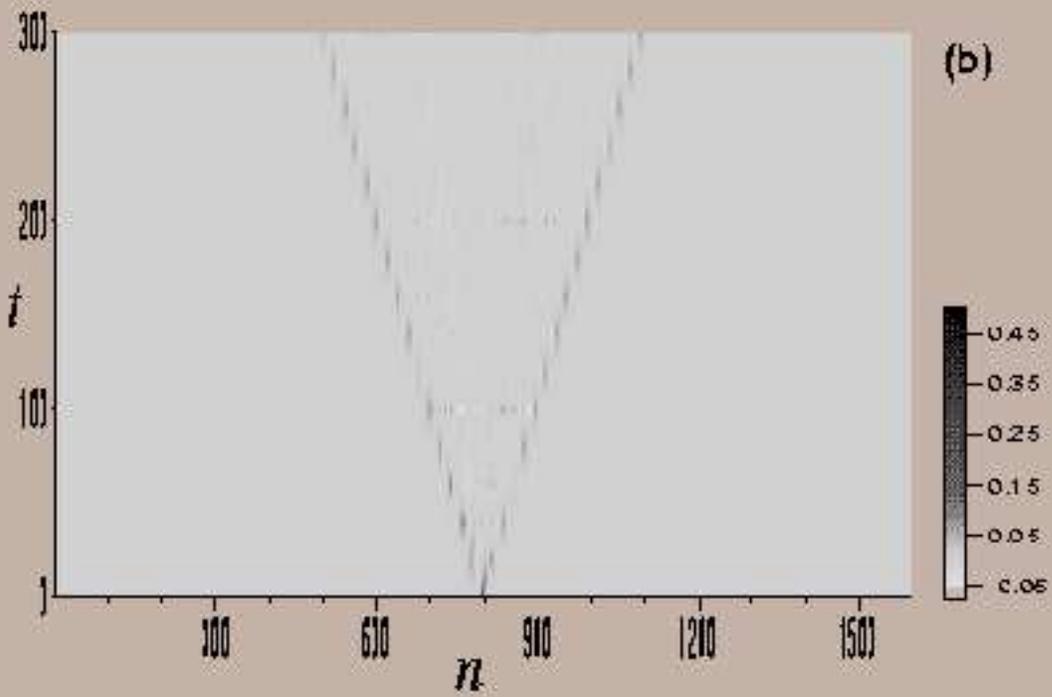
\end{document}